\documentclass{iopart}
\usepackage{graphicx}

\newcommand{\Zeff}{$Z_\mathrm{eff}$}
\newcommand{\eps}{$\epsilon_\mathrm{ff}$}
\newcommand{\dens}{$n_\mathrm{e}$}
\newcommand{\temp}{$T_\mathrm{e}$}

\newcommand{\ud}{\mathrm{d}}

\begin{document}
\title{Reconstruction of \Zeff\ profiles at TEXTOR through Bayesian source separation.}
\author{G. Verdoolaege, G. Telesca and G. Van Oost}
\address{Department of Applied Physics, Ghent University, Rozier 44, 9000 Gent, Belgium}

\begin{abstract}
We describe a work in progress on the reconstruction of radial profiles for the ion effective charge \Zeff\ on the TEXTOR tokamak, using statistical data analysis techniques. We introduce our diagnostic for the measurement of bremsstrahlung emissivity signals. \Zeff\ profiles can be determined by Abel-inversion of line-integrated measurements of the bremsstrahlung emissivity (\eps) from the plasma and the plasma electron density (\dens) and temperature (\temp). However, at the plasma edge only estimated values are routinely used for \dens\ and \temp, which are moreover determined at different toroidal locations. These various uncertainties hinder the interpretation of a \Zeff\ profile outside the central plasma. \\
In order to circumvent this problem, we propose several scenarios meant to allow the extraction by (Bayesian) Blind Source Separation techniques of either (line-integrated) \Zeff\ waveshapes or absolutely calibrated signals from (line-integrated) emissivity signals, using also density and temperature signals, as required.
\end{abstract}

\section{The role of impurities and \Zeff\ in a tokamak plasma}
The behavior of plasma impurities in a tokamak plasma is a critical issue~\cite{Wesson}. On the one hand, these impurities increase the fuel dilution and, at reactor parameters, they can be responsible for a considerable power loss from the plasma core by bremsstrahlung. On the other hand, line radiation by low-$Z$ impurities at the plasma edge can be advantageous for the reduction of the plasma-wall interaction. \\
To study the behavior of impurities in the plasma, one needs a local measure for impurity concentration. A parameter of great interest is the so-called ion effective charge of the plasma, \Zeff, which is averaged over all species. Once the plasma electron density (\dens) and temperature (\temp) are known, it is proportional to the bremsstrahlung emissivity (\eps) from the plasma~\cite{Schoon}, which originates mainly from the acceleration of electrons in the field of ions:

\begin{equation} \label{eq:Zeff}
\epsilon_\mathrm{ff} \sim \frac{n_\mathrm{e}^2 Z_\mathrm{eff}}{\sqrt{T_\mathrm{e}}}, \quad \mathrm{where} \quad
Z_\mathrm{eff} \equiv \frac{\sum_i n_i Z_i^2}{\sum_i n_i Z_i},
\end{equation} 

\noindent and where $n_i$ and $Z_i$ are the density and charge of ion species $i$.

\section{A diagnostic for the determination of \Zeff\ on TEXTOR}
\begin{figure}
\begin{center}
\includegraphics[height=7cm]{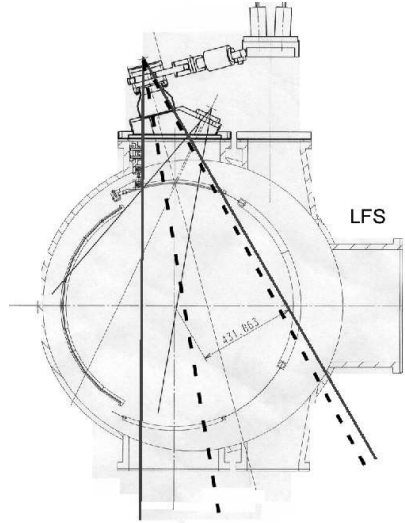}
\end{center}
\caption{\label{fig:Geometry} The viewing geometry for the bremsstrahlung \Zeff\ diagnostic at TEXTOR. The dashed lines represent two of the viewing chords used in the ICA analysis described below.}
\end{figure}

At the tokamak TEXTOR (FZJ J\"ulich, Germany), we run a diagnostic for the determination of \Zeff\ from measurements of bremsstrahlung emissivity in the visible (see also \cite{Schoon}). 21 fiber optic cables are directed along different lines of sight in a poloidal cross-section of the plasma (figure \ref{fig:Geometry}). The light is led through an interference filter (peak transmission at 523 nm) and focused onto the CCD of a cooled 12-bit camera (DTA Inc., Italy). The transmitted wavelength region is known to consist for 99\% of bremsstrahlung continuum. The measured emissivity signals are stored into the internet-accessible TEXTOR central database. \\
In addition, four other channels will be equipped with a photomultiplier, in order to study some phenomena taking place on smaller time scales as well. We expect to have this second system operational by the beginning of the TEXTOR campaign starting in the last quarter of 2004.

\begin{figure}  
\begin{center}
\includegraphics[height=7cm]{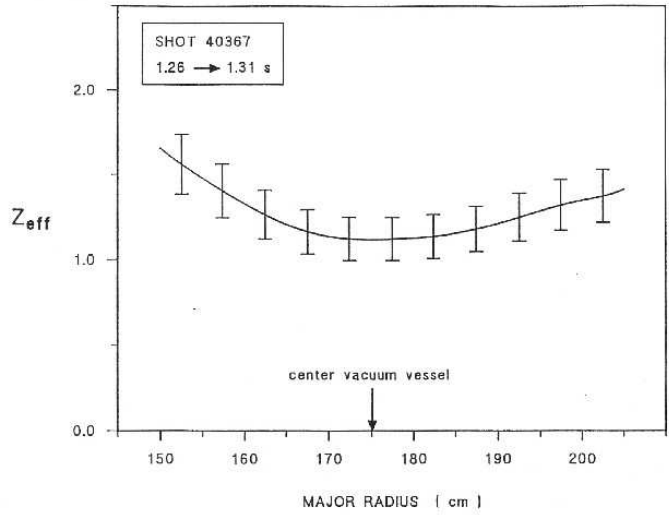}
\end{center}
\caption{\label{fig:Profile} A typical \Zeff\ profile obtained by Abel-inversion of line-integrated quantities. The error bars are obtained from the estimated errors on the emissivity, density and temperature profiles, as described in \cite{Schoon}.}
\end{figure}

\noindent A central line-averaged \Zeff\ is routinely computed and is available online. The average \Zeff\ is defined as (see \cite{Foord})

\begin{equation} \nonumber
\overline{Z}_\mathrm{eff} \sim \frac{\int_{-a}^a \epsilon(r)\ud r}{\int_{-a}^a \overline{g}_\mathrm{ff}n_\mathrm{e}^2(r) T_\mathrm{e}^{-1/2}(r) \ud r},
\end{equation} 

\noindent where $a$ is the plasma radius and $\overline{g}_\mathrm{ff}$ the so-called (averaged) Gaunt factor. In addition, a radial profile for \Zeff\ (figure \ref{fig:Profile}) can be reconstructed from the full set of line-integrated emissivities by an Abel-inversion~\cite{Schoon}. This also requires profiles for electron density and temperature.

\section{Difficulties in the reconstruction of \Zeff\ profiles}
\begin{figure} 
\begin{center}
\includegraphics[height=7cm]{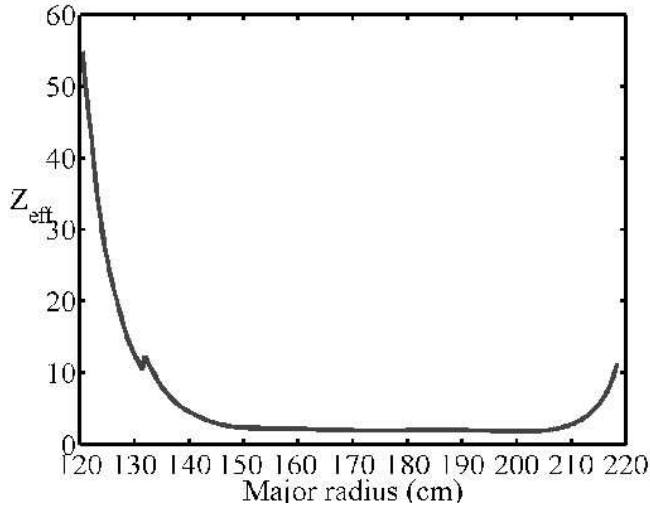}
\end{center}
\caption{\label{fig:BadProfile} An extreme case of the problems that might arise in the interpretation of a reconstructed \Zeff\ profile. The profile diverges at the inboard side of the tokamak.}
\end{figure}

Several factors render the radial matching of the different profiles problematic, hindering the interpretation of a \Zeff\ profile outside the central plasma. Indeed, at the plasma edge only estimated values are routinely used for \dens\ and \temp, which are moreover determined at different toroidal locations. As a consequence, one frequently observes an unphysical rise of \Zeff\ toward the plasma edge (figure \ref{fig:BadProfile}). In fact, so far none of the available methods for the determination of \Zeff\ has provided a full profile, which is at present a real challenge.

\noindent Conversely, if it were e.g. possible to obtain a set of line-integrated values for \Zeff\ directly from the line-integrated measurements of \eps, \dens\ and \temp, then these problems would be avoided, as a \Zeff\ profile could directly be obtained from inversion of the line-integrals. Alternatively, one may work directly with already inverted data. As explained below, this even opens the possibility to obtain estimations for \Zeff\ without using density and temperature information. \\
In a first attempt toward these goals, we want to get an idea of the relative signal of \Zeff\, using bremsstrahlung emissivity measurements and if necessary also density and temperature signals.

\section{Extraction of a \Zeff\ signal from line-integrated bremsstrahlung measurements; test using linear multi-channel ICA}
\begin{figure} 
\begin{center}
\includegraphics[height=7cm]{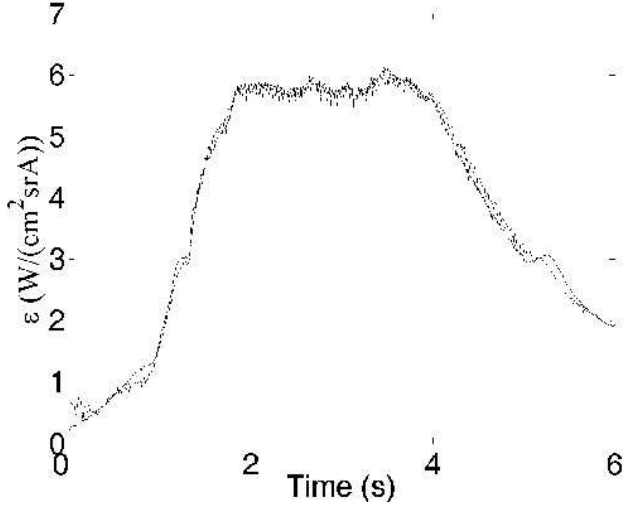}
\end{center}
\caption{\label{fig:Regression} A line-integrated emissivity signal (solid curve) and its fit using a linear regression of a line-integrated \eps\ on line-integrals (same line of sight) of $n_\mathrm{e}^2$ and \Zeff. Despite of the several simplifications, the regression model (\ref{eq:Regression}) seems to hold.}
\end{figure}

From equation (\ref{eq:Zeff}), the signal for \Zeff\ can be seen as a nonlinear mixture of the signals for \eps, \dens\ and \temp. The dependence on \temp\ is, however, weak and the \temp-dependence can in a first approximation be left out of the analysis. We may linearize the expression of \eps\ as a function of $n_\mathrm{e}^2$ and \Zeff\ up to first order, i.e.:

\begin{equation} \label{eq:Regression}
\epsilon_\mathrm{ff} \approx C_1 n_\mathrm{e} + C_2 Z_\mathrm{eff}, \quad \textrm{$C_1$, $C_2$ regression coefficients,}
\end{equation} 

\noindent which constitutes a linearized mixing model. The traditionally obtained signals for the values of the plasma parameters can be used to perform a linear regression analysis, and a similar analysis can be performed for the corresponding line-integrated signals, upon radial integration of the model (\ref{eq:Regression})\footnote{In fact, one finds that one may even use directly \dens\ (and \temp) as input signals for the regression.}. The results are displayed in figure \ref{fig:Regression}.

\begin{figure} 
\begin{center}
\includegraphics[height=7cm]{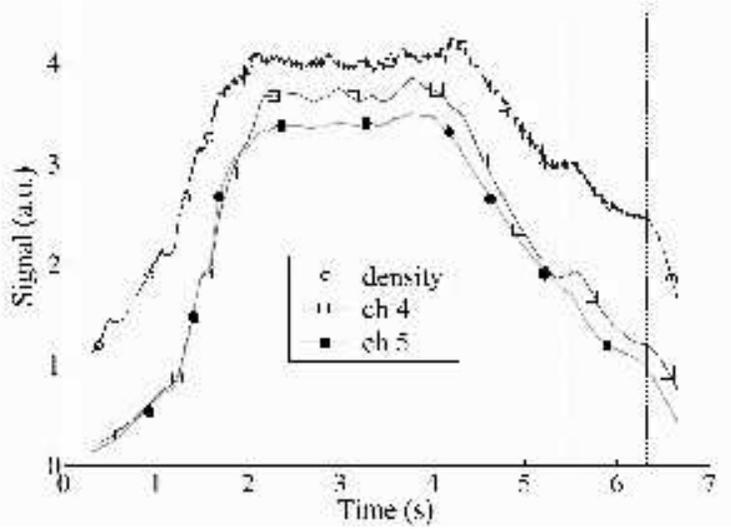}
\end{center}
\caption{\label{fig:EPS1} The time trace of the line-integrated emissivity from a central channel (ch 4) and a more peripheral channel (ch 5) (shown in figure \ref{fig:Geometry}). The signal (rescaled) of the central line-integrated density is also shown. Two striking features in the signal are marked by dashed lines.}
\end{figure}

\begin{figure} 
\begin{center}
\includegraphics[height=7cm]{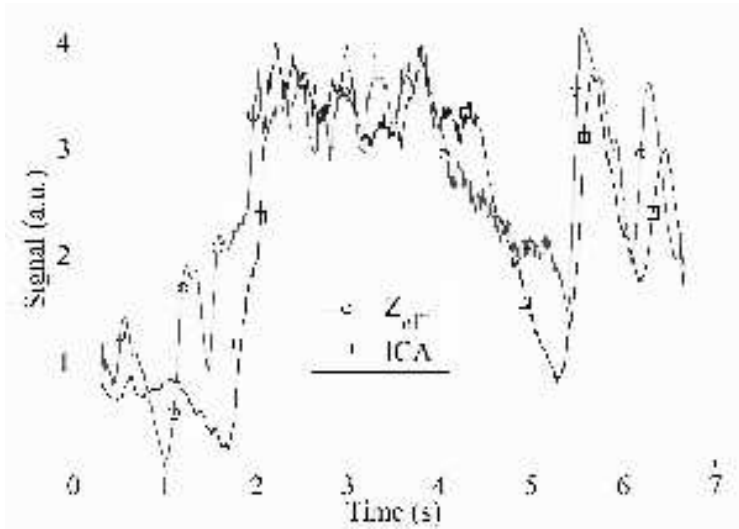}
\end{center}
\caption{\label{fig:EPS2} The time trace (ICA) of one of the extracted independent components and the signal (rescaled) of the central \Zeff. Note the good correspondence, especially at the time of the bumps in the emissivity signal.}
\end{figure}

\noindent As a preliminary test (see also \cite{EPS} and \cite{FLINS}) for the process of extraction of a \Zeff\ signal from bremsstrahlung emissivity signals, four emissivity line-integrals along different lines of sight were fed to the FastICA program \cite{FastICA} for linear Independent Component Analysis (ICA) \cite{ICA}. \\
ICA is a technique that can be derived in a Bayesian context from first principles. In (linear) ICA, it is assumed that a number $n$ of signals $x_i(t)$ is measured, which are each a linear mixture of (ideally) the same number of unknown source signals $s_i(t)$ that are assumed to be statistically independent:

\begin{equation}
\left[
\begin{array}{c}
x_1(t) \\
x_2(t) \\
\vdots \\
x_n(t)
\end{array}
\right]
=
\mathbf{A}
\left[
\begin{array}{c}
s_1(t) \\
s_2(t) \\
\vdots \\
s_n(t)
\end{array}
\right],
\end{equation} 

\noindent where $\mathbf{A}$ is an unknown mixing matrix. ICA now consists of estimating both the source signals and the mixing matrix. The idea is in principle to find a set of linear combinations of the $x_i$ that are maximally nongaussian. Due to the central limit theorem these will equal the independent components $s_i$. Thus, the ICA procedure requires the maximization  of some measure of nongaussianity (e.g. negentropy), which is accomplished by some gradient descent approach. \\
In our application, it is thus implicitly assumed that the waveform for any time trace is the same at different radii, which is on the average true, and any local diversion from this average will be considered as due to noise, or is neglected while taking different regression coefficients for different lines of sight. The results of this analysis are displayed in figures \ref{fig:EPS1} and \ref{fig:EPS2}, and can be compared to the central signal for \Zeff\ (figure \ref{fig:EPS2}). In this analysis, no density or temperature information was used.

\noindent It should be noted however that the signals for the plasma parameters are, to a certain extent, interrelated, and thus the assumption of independence of the components is not entirely fulfilled in this instance. Also, the obtained \Zeff\ waveform is in fact an average over a certain viewing angle in the poloidal cross-section, and can thus hardly be used for the reconstruction of a reliable \Zeff\ profile. These issues are treated in the next section.

\section{Outlook: Bayesian source separation models}
The signals for \eps, \dens, \temp\ and \Zeff\ are mutually dependent, so the independence assumption of the ICA scheme does not hold, although in practice it still yields acceptable results, as demonstrated above. We would however like to have more control over the actual source signal that is extracted from the mixture, in order to obtain better estimations of a \Zeff\ waveform during discharges in different plasma regimes. \\
Here, the Bayesian line of thought might come to help, as it allows us to incorporate prior information on the physics of the problem and the \Zeff\ waveshape into the separation model. This information can be extracted from the central and line-averaged \Zeff\ signals obtained via traditional methods. \\
In practice, the aim in Bayesian source separation is to obtain the posterior distribution of the unknown signals $S$ in terms of the likelihood of the measured data $X$ and the prior information (encoded in $p(S)$) one is willing to assume: 

\begin{equation}
p(S|X) \sim p(X|S)p(S).
\end{equation} 

\noindent We are thus looking for signals $S$ that maximize this posterior probability, given the prior distribution and the measured data. We then also use known theoretical relations between the data and the signals $S$, such as (\ref{eq:Zeff}). \\
This analysis can be conducted as a single-channel method (see \cite{Knuth}) on line-integrated signals, and when it is done for several lines of sight, it allows us to directly obtain a \Zeff\ profile via Abel-inversion. As mentioned before, instead one can also operate immediately on profile data.

\section{Conclusion}
Several difficulties are associated with the calculation of \Zeff\ profiles in a tokamak plasma, in particular at the plasma edge. Determining \Zeff\ signals directly from the bremsstrahlung data can avoid these problems. We have conducted a preliminary test with good results for the construction of \Zeff\ line-integrals via Independent Component Analysis. Finally, we have argued that the separation process can be made more robust using a Bayesian source separation technique, which can employ prior information on the \Zeff\ signal.

\section*{Acknowledgments}
This work is partially funded by the NSF Flanders, Belgium. We would like to acknowledge its support.

\end{document}